\documentclass[11pt]{article}
\usepackage{graphicx,amsmath,amsthm,mathrsfs,amssymb}
\usepackage[usenames]{color}
\usepackage{ulem}
\usepackage{authblk}

\newcommand{\ee}{\end{equation}}
\newcommand{\word}[1]{\,\,\mbox{#1}\,\,}
\newcommand{\reff}[1]{(\ref{#1})}
\newcommand{\beq}{\begin{equation}}
\newcommand{\eeq}[1]{\label{#1}\end{equation}}
\newcommand{\beqa}{\begin{eqnarray}}
\newcommand{\eea}{\end{eqnarray}}
\newcommand{\eeqa}[1]{\label{#1}\end{eqnarray}}
\newcommand{\beg}{\begin{equation*}}
\newcommand{\eeg}{\end{equation*}}

\newcommand{\bsplit}{\begin{split}}
\newcommand{\esplit}{\end{split}}

\title{Restricted Weyl invariance in four-dimensional curved spacetime}

\author[1]{Ariel Edery\thanks{aedery@ubishops.ca}}
\author[2]{Yu Nakayama\thanks{nakayama@theory.caltech.edu}}

\affil[1]{Department of Physics, Bishop's University, \\2600 College Street, Sherbrooke, Qu\'{e}bec, Canada, J1M 1Z7 .\vspace{8mm}}
\affil[2]{Kavli Institute for the Physics and Mathematics of the Universe (WPI), Todai Institutes for Advanced Study, University of Tokyo,
5-1-5 Kashiwanoha, Kashiwa, Chiba 277-8583, Japan.}
\begin{document}
\date{}
\maketitle
\begin{abstract}
We discuss the physics of {\it restricted Weyl invariance}, a symmetry of dimensionless actions in four dimensional curved space time. 
When we study a scalar field nonminimally coupled to gravity with Weyl(conformal) weight of $-1$ (i.e. scalar field with the usual two-derivative kinetic term), we find that dimensionless terms are either fully Weyl invariant or are Weyl invariant if the conformal factor $\Omega(x)$ obeys the condition $g^{\mu\nu}\nabla_{\mu}\nabla_{\nu}\Omega=0$. We refer to the latter as {\it restricted Weyl invariance}. We show that all the dimensionless geometric terms such as $R^2$, $R_{\mu\nu}R^{\mu\nu}$ and $R_{\mu\nu\sigma\tau}R^{\mu\nu\sigma\tau}$ are restricted Weyl invariant. Restricted Weyl transformations possesses nice mathematical properties such as the existence of a composition and an inverse in four dimensional space-time. We exemplify the distinction among rigid Weyl invariance, restricted Weyl invariance and the full Weyl invariance in dimensionless actions constructed out of scalar fields and vector fields with Weyl weight zero.

\end{abstract}
\setcounter{page}{1}
\section{Introduction}

An action in curved spacetime is said to possess Weyl symmetry \cite{Weyl} if it is invariant under the following Weyl transformation: $g_{\mu\nu} \to \Omega^2(x) g_{\mu\nu}$ and $\psi \to \Omega^s \psi$ where $g_{\mu\nu}$ is the metric, $\Omega(x)$ is the conformal factor, a continuous non-vanishing function of $x$,  $\psi$ refers to a generic field and $s$ is the Weyl (conformal) weight of the field $\psi$ \footnote{Throughout this paper, a local rescaling of the metric $g_{\mu\nu} \to \Omega^2(x) g_{\mu\nu}$ will be referred to as a Weyl transformation instead of a conformal transformation. However, we place in brackets the word ``conformal" whenever its usage has become common in the literature such as ``conformal to flat"  or the ``conformal weight" of a field.}. A generic dimensionless action in four dimensional curved spacetime will not in general possess Weyl symmetry because not all dimensionless terms are Weyl invariant. For example, if $R$ is the Ricci scalar, such an action can contain dimensionless terms such as $R^2$ which are not Weyl invariant. Nonetheless, as we show, $R^2$ still possesses a large symmetry: it is invariant under Weyl transformations that obey the condition $\Box \Omega\equiv g^{\mu\nu}\nabla_{\mu}\nabla_{\nu}\Omega =0$. We refer to this as {\it restricted Weyl symmetry}.  Restricted Weyl symmetry is a much larger symmetry than rigid Weyl invariance where $\Omega$ must be a constant. In this paper, we start with a study of a dimensionless action where the scalar field $\phi$ has the typical well-behaved two-derivative kinetic term $g^{\mu\nu} \partial_{\mu}\phi \partial_{\nu} \phi$. This is an action where the scalar field has a Weyl (conformal) weight of $-1$.  We show that the minimal kinetic term for $\phi$ (i.e. $g^{\mu\nu} \partial_{\mu}\phi\partial_{\nu}\phi$) as well as the non-minimal coupling term $\eta R \phi^2$ where $\eta$ is an arbitrary dimensionless coupling constant, are also examples of terms which are not Weyl invariant but possess restricted Weyl invariance. Though the dimensionless action with scalar field of Weyl weight $-1$ is not generically Weyl invariant, we show it is nonetheless always restricted Weyl invariant. To the best of our knowledge, restricted Weyl symmetry has not been emphasized much in the physics literature (see \cite{O'Raifeartaigh:1996hf}, however, in which the ``harmonic Weyl group" was mentioned).

Restricted Weyl transformations possess some interesting properties. Two consecutive restricted Weyl transformations form a restricted Weyl transformation by a composition law in four dimensions. As a corollary, restricted Weyl transformations possess an inverse but again only in four dimensions. Restricted Weyl symmetry leads one naturally to consider metrics which are Weyl flat (conformal to flat) but where the conformal factor $\Omega$ obeys $\Box \Omega=0$. We refer to such metrics as {\it restricted Weyl flat} metrics. In a restricted Weyl flat metric, both the Weyl tensor and the Ricci scalar are zero. Well-known spacetimes happen to be restricted Weyl flat. Examples include $AdS_2 \times S^2$ and the radiation-dominated era of the FLRW expanding universe. In particular, particle creation in an expanding FLRW spacetime due to the Weyl-symmetry breaking term $R^2$ that appears in the trace anomaly does not occur in the radiation-dominated era since $R^2$ vanishes.     

For completeness, we discuss actions which are rigid Weyl invariant (but are not necessarily full or restricted Weyl invariant) in four dimensions. This leads us naturally to depart from two-derivative kinetic terms to consider higher-derivative kinetic terms. The simplest case is four derivatives where the scalar field has Weyl weight of zero. We find that the most generic dimensionless action of this type contains terms which are only rigid Weyl invariant besides terms which are restricted Weyl and Weyl invariant. We then discuss a physically relevant model involving vector fields which is rigid Weyl invariant (but is not necessarily full or restricted Weyl invariant) in a Euclidean spacetime. When the vector fields have gauge invariance, the dimensionless terms are all Weyl invariant,   but this is no longer true once the gauge invariance is explicitly broken as in the theory of elasticity.

In the context of the local renormalization group, these dimensionless actions with Weyl weight zero appear as the local part of the ``Schwinger functional" of the quantum field theories, where the Weyl weight zero fields are interpreted as the source functions of the given quantum field theory under consideration \cite{Osborn:1991gm} (see also \cite{Yu} for a review). In particular, in the recent discussions of the proof of the ``a-theorem" in four dimensional space-time \cite{Komargodski:2011vj}, the so-called ``on-shell Weyl condition" $\eta_{\mu\nu} \partial^\mu \partial^\nu \Omega = 0$ in the flat space-time limit for this Schwinger functional was introduced. We may regard our restricted Weyl condition as a curved space-time generalization of it, but we will see that the mathematically interesting structure such as the composition law only appears in our curved space-time generalization.

\section{Restricted Weyl invariance in four dimensions}

Consider a dimensionless action with a scalar field nonminimally coupled to gravity and with the usual kinetic term containing two derivatives. The scalar field therefore has a Weyl weight of $-1$.  The action is necessarily massless and all coupling contants are dimensionless. A generic action of this type is
\beq
S = \int d^4x \sqrt{|g|}\Big( -g^{\mu\nu} \partial_{\mu} \phi \partial_{\nu} \phi - \eta R \phi^2 + a \,\mathrm{G}  -c\,\mathrm{Weyl}^2 + \frac{b}{9}\,R^2 -\dfrac{\lambda\phi^4}{4!} \Big) \,.
\eeq{S1}
where 
\beq
\mathrm{G}\equiv R_{\mu\nu\sigma \tau}R^{\mu\nu\sigma\tau} - 4 R_{\mu\nu}R^{\mu\nu} +R^2 
\eeq{Euler}
is the Gauss-Bonnet toplogical invariant, and 
\beq
\mathrm{Weyl}^2 \equiv C_{\mu\nu\sigma \tau}C^{\mu\nu\sigma\tau}=R_{\mu\nu\sigma \tau}R^{\mu\nu\sigma\tau} - 2 R_{\mu\nu}R^{\mu\nu} +\dfrac{1}{3}\,R^2 
\eeq{Weyl}
is the Weyl tensor squared. Here,
$R$ and $R_{\mu\nu}$ are the Ricci scalar and Ricci tensor respectively. The constants $a, b, c$ and $\eta$ are all dimensionless. 
The above action \reff{S1} was the starting point for a detailed study of the trace anomaly in $\lambda \phi^4$ theory by Hathrell \cite{Hathrell}.  

This action is contrasted with the classically Weyl invariant scalar action (typically known as the ``conformally coupled scalar action")
\beq
S_{\text{conf}} = \int d^4x \sqrt{|g|} \Big(-g^{\mu\nu} \partial_{\mu} \phi \partial_{\nu} \phi  -\frac{1}{6} R \phi^2 - \frac{\lambda \phi^4}{4!}\Big) \,.
\eeq{S2}
which is invariant under the following Weyl transformation
\beq
g_{\mu\nu} \to \Omega^2(x) g_{\mu\nu} \word{and} \phi \to \dfrac{\phi}{\Omega}
\eeq{Transform}
where $\Omega(x)$ is a continuous non-vanishing function of $x$. The scalar field has a Weyl (conformal) weight of $-1$ as in action \reff{S1}.  
As a quantum field theory, the distinction between \reff{S1} and \reff{S2} are obscured because all the higher-derivative curvature terms in \reff{S1} are generated by local quantum corrections to \reff{S2}. Moreover, the renormalization process changes the coefficient $1/6$ of $R \phi^2$ in the action \reff{S2} to an arbitrary constant $\eta$ in \reff{S1}. Unlike \reff{S2}, the action \reff{S1} is not Weyl invariant ($c$ would have to be zero and $\eta$ would have to be $1/6$ for the action to be Weyl invariant). The quantum corrections break the Weyl invariance of the original two-derivative action. One goal of this section is to show that the action \reff{S1} remains invariant under the Weyl transformation \reff{Transform} if the Weyl factor $\Omega(x)$ obeys the condition 
\beq
\Box \,\Omega(x) \equiv g^{\mu\nu}\nabla_{\mu}\nabla_{\nu} \Omega(x)=0\,.
\eeq{restricted}
We call this {\it restricted Weyl invariance}.\footnote{Of course, as a quantum field theory, the (restricted) Weyl invariance is typically broken due to the renormalization process and the trace anomaly. These may be cancelled by further introducing the compensator ``dilaton" and regard \reff{S1} as the local part of the effective action. Our discussions in this paper are, however, essentially classical after we obtain the local effective action such as \reff{S1}.}

Under the Weyl transformation \reff{Transform} the terms $\mathrm{Weyl}^2$, $\mathrm{G}$ and $\lambda \phi^4$ are all invariant\footnote{G is Weyl invariant up to a total derivative \cite{Yu} (see appendix B, Eq. \reff{Euler2}). However, the total derivative makes no contribution to the action.}(when multiplied by the factor $\sqrt{|g|}$). We refer to such terms as Weyl invariant. The other three remaining terms in \reff{S1} which include the kinetic term for $\phi$, the nonminimally coupled term $\eta R \phi^2$ and the Ricci scalar squared $R^2$ are not Weyl invariant by themselves. However, each of the three terms are restricted Weyl invariant as we now show. 

Under the Weyl transformation \reff{Transform} we have the following transformation rules \cite{Parker, BD}
\begin{align}
\sqrt{|g|} &\to \Omega^4 \sqrt{|g|} \label{detg}\\
 R &\to \Omega^{-2}\,R - 6 \,\Omega^{-3}\,\Box \,\Omega\,.
\label{R}\end{align}
It follows then that under a Weyl transformation we have
\begin{align}
\sqrt{|g|} \,R \,\phi^2 &\to \sqrt{|g|} \,R \,\phi^2 - \sqrt{|g|}\, 6 \,\phi^2\,\Omega^{-1}\,\Box \Omega \, \label{Rphi2}\,,\\
\sqrt{|g|} \,R^2&\to\sqrt{|g|} \,R^2 - \sqrt{|g|} \,12 \,R \,\Omega^{-1}\, \Box \Omega + \sqrt{|g|} \,36  \,\Omega^{-2}\,(\Box \, \Omega)^2\,.
\label{R2}\end{align}

Clearly, the above two terms are not Weyl invariant. However, they are invariant if the Weyl factor $\Omega$ obeys $\Box\,\Omega=0$; they are restricted Weyl invariant. That $R^2$ is restricted Weyl invariant might have important consequences for particle production in an expanding FLRW expanding spacetime in the early epoch. We discuss this later.

The minimal kinetic term for the scalar field also possesses restricted Weyl invariance. We calculate how it transforms under a Weyl transformation in appendix A. The result is given by \reff{Kinetic2}
\beq
g^{\mu\nu} \partial_{\mu} \phi \partial_{\nu} \phi \to 
\sqrt{|g|}\, g^{\mu\nu} \partial_{\mu}\phi \partial_{\nu}\phi + \sqrt{|g|}\, \phi^2 \,\Omega^{-1}\,\Box \,\Omega - \sqrt{|g|}\, \nabla_{\mu} \big(\phi^2 \,\nabla^{\mu} (\ln \Omega) \big)\,.
\eeq{Kinetic}
The last term in the above expression is a total derivative and makes no contribution to the action. The existence of the second term implies that the minimal kinetic term is not Weyl invariant. However, if $\Box \Omega =0$ then the minimal kinetic term is invariant; it is restricted Weyl invariant. 

We have therefore shown that the action \reff{S1} is restricted Weyl invariant. The transformation properties of $R \phi^2$ and the minimal kinetic term for 
$\phi$ given by \reff{Rphi2} and \reff{Kinetic} respectively allows one to construct the well known Weyl invariant quantity, the ``conformal kinetic term" $g^{\mu\nu} \partial_{\mu} \phi \partial_{\nu} \phi +\dfrac{1}{6} R\,\phi^2$. The action \reff{S1} is therefore Weyl invariant if one chooses specific coefficients for the terms i.e. $b=0$ and $\eta=1/6$. But for arbitrary constants it possesses only restricted Weyl symmetry. In particular, we found that $R^2$, the minimal kinetic term for $\phi$ (i.e. $g^{\mu\nu} \partial_{\mu}\phi\partial_{\nu}\phi$) and $R \,\phi^2$ all possess restricted Weyl symmetry in four dimensions. 

Of the higher-derivative geometrical terms, the Weyl tensor squared $\mathrm{Weyl}^2$ is Weyl invariant by itself and the Gauss-Bonnet term $\mathrm{G}$ is Weyl invariant up to a total derivative. We have now shown that $R^2$ is restricted Weyl invariant. These three terms form a complete basis of the charge-parity (CP) even dimensionless geometric actions in four dimension because $R_{\mu\nu}R^{\mu\nu}$ and $R_{\mu\nu\sigma \tau}R^{\mu\nu\sigma\tau}$ can be expressed in terms of $\mathrm{Weyl}^2$ , $\mathrm{G}$ and $R^2$ via \reff{Euler} and \reff{Weyl}. For completeness, the transformation properties of $R_{\mu\nu}R^{\mu\nu}$ and $R_{\mu\nu\sigma \tau}R^{\mu\nu\sigma\tau}$ are calculated in appendix B and given by equations \reff{R2R} and \reff{R3R}. Like $R^2$,  both terms are restricted Weyl invariant (up to a total derivative which makes no contribution to the action).

In four space-time dimension, there is one additional CP odd dimensionless geometric term $\epsilon_{\alpha\beta \gamma \delta}R_{\alpha\beta\mu\nu} R^{\mu\nu}_{\ \ \gamma \delta}$ known as the Hirzebruch-Pontryagin density. The term is topological and sometimes called the gravitational $\theta$ term, and it is Weyl invariant. In the most part of this paper, we assume the CP invariance, so we will neglect these CP breaking terms. There are no known unitary quantum field theories in which the gravitational $\theta$ term is physically renormalized \cite{Nakayama:2012gu}.

\subsection{Properties of restricted Weyl transformations}
Let us study some properties of the restricted Weyl transformations.
Consider the two consecutive restricted Weyl transformations in general $d$ space-time dimensions
\begin{align}
\tilde{g}_{\mu\nu} &= \Omega^2 g_{\mu\nu} \cr
\tilde{\tilde{g}}_{\mu\nu} & = \tilde{\Omega}^2 \tilde{g}_{\mu\nu} = \tilde{\Omega}^2 \Omega^2 g_{\mu\nu} \cr
\end{align}
where $\Box \Omega = 0$ and $\tilde{\Box} \tilde{\Omega} = 0$.
The question is whether the consecutive restricted Weyl transformation satisfies the composition law such that $\Box (\tilde{\Omega}\Omega) = 0$. 

First of all $\tilde{\Box} \tilde{\Omega} = 0$ means
\begin{align}
\Omega^{-2} \Box  \tilde{\Omega} + (d-2)\Omega^{-3}g^{\mu\nu} \partial_\mu \Omega \partial_\nu \tilde{\Omega} = 0 \ .
\end{align}
Thus,
\begin{align}
\Box (\tilde{\Omega}\Omega) &=\Omega \Box \tilde{\Omega} + 2g^{\mu\nu}\partial_\mu \tilde{\Omega} \partial_\nu \Omega + \tilde{\Omega} \Box \Omega \cr
 & = -(d-2) g^{\mu\nu}\partial_\mu \Omega \partial_\nu \tilde{\Omega} + 2g^{\mu\nu} \partial_\mu \tilde{\Omega} \partial_\nu \Omega
\end{align}
so that in (and only in) $d=4$ dimension, the consecutive restricted Weyl transformation generates the restricted Weyl transformation by a composition law.\footnote{In $d\neq 4$ space-time dimensions, to retain the composition law, one has to impose the non-linear condition $\Omega \Box\Omega + \frac{d-4}{2}g^{\mu\nu}\partial_\mu \Omega\partial_\nu\Omega = 0$. See also \cite{O'Raifeartaigh:1996hf}.} Note that ``consecutive" means that the d'Alembertian must be evaluated for the Weyl transformed metric $\tilde{\Box} \tilde{\Omega} = 0$ rather than $\Box \tilde{\Omega} = 0$.

The ``inverse" of the restricted Weyl transformation exists in $d=4$ dimensions only. For
\begin{align}
\tilde{g}_{\mu\nu} &= \Omega^2 g_{\mu\nu} \cr
\end{align}
with $\Box \Omega = 0$, we may define
\begin{align}
g_{\mu\nu}  = \Omega^{-2} \tilde{g}_{\mu\nu} \ 
\end{align}
as an inverse of the restricted Weyl transformation. It can be readily checked that $\tilde{\Box} \Omega^{-1} = 0$ only in four dimensions.

We would like to stress that the multiplication with respect to ``independent" restricted Weyl transformations $\Omega_1$ and $\Omega_2$ with $\Box \Omega_1 = \Box \Omega_2 = 0$ does not form the multiplicative group $\Omega_{1\times 2} = \Omega_1 \Omega_2$ (unlike in the unrestricted case) because $\Box (\Omega_1 \Omega_2) \neq 0 $ in general. On the other hand, the ``addition" forms an Abelian group $\Omega_{1+2} = \Omega_1 + \Omega_2$ because of the linearity of the condition. This Abelian group structure is valid in any space-time dimension.

\subsection{Restricted Weyl flat metrics}

Let $\eta_{\mu\nu}$ represent the flat Minkowski metric. If a metric $g_{\mu\nu}$ is Weyl flat (conformal to flat) so that $g_{\mu\nu} = \Omega^2(x) \eta_{\mu\nu}$, then the Weyl tensor is zero. If $\Omega$ obeys $\Box \Omega=0$, where $\Box$ here is the flat space d'Alembertian,  then the Ricci scalar is also zero; this stems immediately from the transformation property \reff{R} of the Ricci scalar and the fact that $R=0$ in Minkowski spacetime.  We will call a Weyl flat metric which obeys $\Box \Omega=\eta^{\mu\nu}\partial_{\mu}\partial_{\nu} \Omega=0$, a {\it restricted Weyl flat metric}.  Both the Weyl tensor and the Ricci scalar are zero in a restricted Weyl flat metric. Moreover, from \reff{Weyl}  we see that the Ricci tensor cannot be zero in a restricted Weyl flat spacetime (except for the trivial case of flat spacetime i.e. if $\Omega$ is a constant). In the context of General Relativity, restricted Weyl flat spacetimes are never vacuum spacetimes but are spacetimes with traceless matter since $R=0$. 

Well-known curved spacetimes in four dimensions happen to be restricted Weyl flat. The spacetime $AdS_2 \times S^2$ (i.e. two-dimensional anti-de Sitter space times a two-sphere) which represents the near horizon limit of an extremal black hole and the radiation dominated era of FLRW cosmology are both examples of restricted Weyl flat spacetimes. The metric of $AdS_2 \times S^2$ is given by
\beq
ds^2=-\dfrac{dt^2}{r^2} + \dfrac{dr^2}{r^2} + (d\theta^2 + \sin^2 \theta d\phi^2)=\dfrac{1}{r^2} \big(-dt^2 +dr^2 + r^2 (d\theta^2 + \sin^2 \theta d\phi^2) \big)\,.
\eeq{AdS2}
It is Weyl flat with Weyl (conformal) factor $\Omega=1/r$. Since $\Box \Omega =0$, this is a restricted Weyl flat metric\footnote{Strictly speaking, $\Box \Omega=0$ except at $r=0$.}. One can readily check that metric \reff{AdS2} has a Weyl tensor and a Ricci scalar of zero. 

We will now determine the scale factor $a(t)$ for a flat space FLRW metric that corresponds to a restricted Weyl flat metric. The flat space FLRW metric can be expressed in the following equivalent forms
\beq
ds^2= -dt^2 + a^2(t) (dx^2 +dy^2 +dz^2) = \Omega^2(\tau)(-d\tau^2 + dx^2 +dy^2 +dz^2)
\eeq{FLRW}
where $a(t)$ is the scale factor and $\Omega(\tau)$ is the conformal factor. Solving $\Box \Omega(\tau)=\eta^{\mu\nu}\partial_{\mu}\partial_{\nu} \Omega(\tau)=0$ yields that $\Omega(\tau)= a\,\tau +b =a\,\tau'$ where $a$ and $b$ are constants and $\tau'=\tau + b/a$. The relation between $t$ and $\tau'$ is given by $dt^2=\Omega^2(\tau)d\tau^2=\Omega^2(\tau')d\tau'^2$ so that $t =c\,\tau'^2$ where $c$ is a constant (we set $t=0$ when $\tau'=0$.) The scale factor $a(t)$ is therefore proportional to $t^{1/2}$. This corresponds to the radiation-dominated era \cite{Carroll2}. Therefore the flat space FLRW metric corresponding to the radiation-dominated era is a restricted Weyl flat metric. One can readily check that the metric \reff{FLRW} with $a(t) \propto t^{1/2}$ has a Ricci scalar of zero and a Weyl tensor of zero.

Particle creation in an expanding FLRW spacetime in $\lambda \phi^4$ theory was studied many years ago \cite{BD2}. One conclusion was that the appearance of the Weyl-symmetry breaking term $R^2$ in the trace anomaly $\langle T^{\mu}_{\mu} \rangle$ leads to particle creation. Though $R^2$ breaks Weyl symmetry, we have shown that it still possesses an important symmetry: restricted Weyl symmetry. In a restricted Weyl flat metric, this term is zero. Therefore, in the radiation-dominated epoch of FLRW, particle creation due to the anomalous $R^2$ term does not occur (because $R=0$ during that era and the contribution from $R^2$ to $\langle T^{\mu}_{\mu} \rangle$ is zero during that epoch.)    

\subsection{Solution of $\Box \Omega = 0$ for compact manifold with Euclidean signature}
In a pseudo-Riemmanian manifold, there is an infinite number of functions of $x$ that satisfy $\Box \Omega(x) =0$. In contrast, for a compact Riemannian manifold $\mathcal{M}$ with the Euclidean signature, there is a mathematical theorem that says the solution of $\Box \Omega =0$ is only $\Omega = \text{const}$. To see this, suppose $\Box \Omega = 0$ and consider the energy functional
\begin{align}
E[\Omega]= \int_{\mathcal{M}} \sqrt{|g|} g^{\mu\nu} \partial_\mu \Omega \partial_\nu \Omega = - \int_{\mathcal{M}} \sqrt{|g|} \Omega \Box \Omega = 0.
\end{align}
The first equality is the integration by part on the compact manifold and the second equality is our assumption that $\Box \Omega = 0$. However, due to the positivity of the metric, the left hand side vanishes if and only if $\partial_\mu \Omega =0$, so the restricted Weyl condition $\Box \Omega = 0$ implies the rigid Weyl transformation $\Omega = \mathrm{const}$.

We emphasize that the above argument only works for a compact manifold with the Euclidean signature. In previous sections, we saw non-trivial and physically relevant examples of restricted Weyl transformations in pseudo-Riemannian manifolds.

\section{Rigid Weyl invariance in four dimensions}

The restricted Weyl invariance of the dimensionless action \reff{S1} in four spacetime dimensions relies on the fact that the kinetic term for the scalar field has two derivatives i.e. the scalar field has a Weyl weight of $-1$. If one considers dimensionless actions in four dimensions with higher derivative interactions (where the scalar field no longer has a Weyl weight of $-1$), the action becomes generically only rigid Weyl invariant, where the conformal factor $\Omega(x)$ must be a constant. The most obvious extension to a two-derivative kinetic term is one with four derivatives (the scalar field then has a Weyl weight of zero).  We therefore study the most generic dimensionless action in four dimensions with scalar field of Weyl weight zero\footnote{It is mathematically known that there are no higher derivative Weyl invariant kinetic terms for a scalar field with positive Weyl weight \cite{FG, GG, G}.}. We will see that all three types of terms exist in such an action: Weyl, restricted Weyl and rigid Weyl invariant terms. Therefore, overall, the action is only rigid Weyl invariant for arbitrary coefficients. However, as with action \reff{S1},  by a judicious choice of the coefficients, one can construct Weyl invariant or restricted Weyl invariant combinations. 

Another motivation to study the theories with Weyl weight zero fields in four-dimensional spacetime is their importance in local renormalization group analysis, where we interpret the classical Weyl weight zero fields as the spacetime dependent sources of the Schwinger functional. In general, the Schwinger functional contains non-local terms in addition to the local terms that we have discussed in the effective action. In this context, the local actions we study in this paper is related to the local counterterms one can always add in the renormalized Schwinger functional. For the consistency of the local renormalization group, the study of the (restricted) Weyl transformation has been crucial rather than the rigid Weyl transformation, in particular in relation to the ``a-theorem" in the flat space-time limit. Here we do not take the flat space-time limit to discuss the distinctions between Weyl invariance, restricted Weyl invariance and rigid Weyl invariance.

\subsection{Dimensionless action in four dimensions with scalar field of Weyl weight zero }

We now extend our study to the case where the scalar fields $\phi^i$ have a Weyl weight of zero. This yields a four-derivative action and its most general form (assuming CP invariance) is  
\begin{align}
S  &= \int d^4 x \sqrt{|g|} \Big\{a(\phi) \mathrm{G} - c(\phi) \mathrm{Weyl}^2  + \frac{1}{9}b(\phi) R^2  + \cr
 & + \frac{1}{3}e_{i}(\phi) \partial_\mu \phi^i \partial^\mu R + \frac{1}{6} f_{ij}(\phi) \partial_\mu \phi^i \partial^\mu \phi^j R + \frac{1}{2}h_{ij}(\phi) \partial_\mu \phi^i \partial_\nu \phi^j G^{\mu\nu} \cr
& + \frac{1}{2} a_{ij}(\phi) \Box \phi^i \Box \phi^j + \frac{1}{2} b_{ijk}(\phi) \partial_\mu \phi^i \partial^\mu \phi^j \Box \phi^k + c_{ijkl}(\phi) \partial_\mu \phi^i \partial^\mu \phi^j \partial_\nu \phi^k \partial^\nu \phi^l \Big\} .
\label{S}\end{align}
Here $f_{ij} = f_{ji}$, $h_{ij} = h_{ji}$, $a_{ij} = a_{ji}$, $b_{ijk} = b_{(ij)k} = \frac{1}{2} (b_{ijk} + b_{jki})$, and $c_{ijkl} = c_{(ij)(kl)} = c_{(kl)(ij)}$. We define the Einstein tensor $G_{\mu\nu} = R_{\mu\nu} - \frac{R}{2} g_{\mu\nu}$ such that $\nabla^\mu G_{\mu\nu} = 0$. Similar actions have been considered in a conformally invariant context \cite{Sergei1, Sergei2, Minas1}.  

For an infinitesimal Weyl variation $\delta g_{\mu\nu} = -2\sigma(x) g_{\mu\nu}$ (i.e. $\Omega(x)=1-\sigma(x)$) 
the variation of the action is given by (see appendix C)
\begin{align}
\delta S= \int d^4x \sqrt{|g|} \Big\{ 4 \,a(\phi)\nabla^{\mu}(R \partial_{\mu}\sigma &- 2 R_{\mu}^{\nu} \partial_{\nu}\sigma) + \dfrac{4}{3} b(\phi) R \Box \sigma 
+ \dfrac{1}{3} e_i(\phi) \partial_{\mu}\phi^i(2R \partial^{\mu}\sigma + 6\partial^{\mu} (\Box \sigma) \cr
&+ f_{ij}(\phi)  \partial_{\mu}\phi^i \partial^{\mu}\phi^j \Box \sigma +h_{ij}(\phi)\partial_{\mu}\phi^i \partial_{\nu}\phi^j (\nabla^{\mu}\nabla^{\nu} \sigma - g^{\mu\nu} \Box \sigma) \cr
&-2 a_{ij}(\phi) \,\partial_{\mu}\sigma \partial^{\mu}\phi^i \Box \phi^j-b_{ijk}  \partial_{\mu}\phi^i \partial^{\mu}\phi^j \partial_{\alpha}\sigma \partial^{\alpha}\phi^k \Big\}\,.
\label{DeltaS}\end{align}


The variation of action \reff{S} given by \reff{DeltaS} can be expressed in the following convenient form after integration by parts, 
\begin{align}
\delta S & = - \int d^4x \sqrt{g} (\partial^\mu \sigma) Z_\mu \cr
Z_\mu & = G_{\mu\nu} w_i(\phi) \partial^\nu \phi^i + \frac{1}{3}\partial_\mu (d(\phi)R)  + \frac{1}{3} R Y_i(\phi)\partial_\mu \phi^i + \cr
& + \partial_\mu (U_i(\phi) \Box \phi^i + \frac{1}{2} V_{ij}(\phi) \partial_\nu \phi^i \partial^\nu \phi^j) + S_{ij}(\phi) \partial_\mu \phi^i \Box \phi^j + \frac{1}{2}T_{ijk} \partial_\nu \phi^i \partial^\nu \phi^j \partial_\mu \phi^k \ ,
\label{DeltaS2}\end{align}
where
\begin{align}
w_i(\phi) & = - 8 \partial_i a(\phi) \cr
d(\phi) & = 4b(\phi) \cr
U_i(\phi) & = -2 e_i(\phi) \cr
Y_i(\phi) &= -2e_i(\phi) \cr
V_{ij}(\phi) &= -4 e_{(i,j)}(\phi) + 2f_{ij}(\phi) - h_{ij}(\phi) \cr
 &= -2 (\partial_i e_j (\phi) + \partial_j e_i(\phi) ) + 2f_{ij}(\phi) - h_{ij}(\phi)  \cr
S_{ij}(\phi) &= h_{ij}(\phi) + 2a_{ij}(\phi) \cr
T_{ijk}(\phi) &= 2h_{k(i,j)}(\phi) - h_{ij,k}(\phi) + 2b_{ijk}(\phi) \cr
& = \partial_j h_{ki}(\phi) + \partial_i h_{kj}(\phi) - \partial_k h_{ij}(\phi) + 2b_{ijk}(\phi).
\end{align}

Thus, the infinitesimal Weyl invariance requires
\begin{align}
\partial_i a(\phi) &= 0 \cr
b(\phi) & = 0 \cr
e_i(\phi) & = 0 \cr
f_{ij}(\phi) & = \frac{1}{2} h_{ij}(\phi) \cr
a_{ij}(\phi) & = -\frac{1}{2} h_{ij}(\phi) \cr
b_{ijk}(\phi) & = -h_{k(i,j)}(\phi) +\frac{1}{2}h_{ij,k}(\phi) \cr \label{relats}
\end{align}
while $c(\phi)$, $h_{ij}(\phi)$, $c_{ijkl}(\phi)$ are arbitrary. Therefore, if the coefficients are chosen as above the action \reff{S} is Weyl invariant.

Clearly, the $\mathrm{Weyl}^2$ term with  $c(\phi)$ coefficient and the four-derivative interaction term with $c_{ijkl}(\phi)$ coefficient are Weyl invariant by themselves because these are terms whose coefficients do not appear in \reff{DeltaS}. Similarly, if the term with $a(\phi)$ coefficient does not depend on $\phi^i$, it becomes simply the Euler term and it is Weyl invariant (up to total derivatives). Probably the most non-trivial Weyl invariant is the term from $h_{ij}$ combined with the other terms related by \reff{relats}. These terms are combined into a generalization of the Riegert action $ \int d^4x{\sqrt{|g|}}\phi(\Box^2 + 2G_{\mu\nu} \nabla^\mu \nabla^\nu + \frac{1}{3}\nabla^\mu R \nabla_\mu)\phi$ with the reparametrization invariance of the field theory space (e.g. $h_{ij}$ is ``metric" on the field theory space  $\phi^i$ and $b_{ijk}$ are ``connections" ).\footnote{The Riegert action was introduced as a four-derivative  Weyl invariant scalar action in four space-time dimensions \cite{Riegert:1984kt} (see also \cite{Fradkin:1983tg,PA}). A generalization of the Riegert action for a particular target space was e.g. studied in \cite{Osborn:2003vk} as an $SL(2,\bf{R})$ invariant Schwinger functional for $\mathcal{N}=4$ super Yang-Mills theory.}

For the infinitesimal restricted Weyl invariance (i.e. $\Box\sigma = 0$), the conditions $d(\phi) = U_i(\phi) = V_{ij}(\phi) $ are not imposed so that we only require
\begin{align}
\partial_i a(\phi) &= 0 \cr
e_i(\phi) & = 0 \cr
a_{ij}(\phi) & = -\frac{1}{2} h_{ij}(\phi) \cr
b_{ijk}(\phi) & = -h_{k(i,j)}(\phi) +\frac{1}{2}h_{ij,k}(\phi) \ \label{relat}
\end{align}
while $c(\phi)$, $b(\phi)$ and $f_{ij}(\phi)$, $h_{ij}(\phi)$, $c_{ijkl}(\phi)$ are arbitrary. If the coefficients are chosen as above, then the action \reff{S} is restricted Weyl invariant.

In general, we may have to check the invariance under the finite restricted Weyl transformation to state the restricted Weyl invariance because the restricted Weyl transformation is not a global Lie group symmetry. However, thanks to the composition properties in four space-time dimensions, we do not have to: the terms with the constraint \reff{relat} are restricted Weyl invariant for a finite transformation. In any way, we can explicitly check the finite restricted Weyl invariance by noting the inhomogeneous part of Weyl transformations for $R$ vanishes when $\Box \Omega = 0$, so the $R^2$ term with $b(\phi)$ coefficient and the $R$ term with $f_{ij}(\phi)$ coefficient are restricted Weyl invariant by themselves.

As already stated, for arbitrary coefficients, the action \reff{S} is only rigid Weyl invariant. This is clear from expression \reff{DeltaS2} for the variation of the action. The factor $\partial^{\mu} \sigma$ in \reff{DeltaS2} renders the variation zero only if $\sigma$ is a constant.

\subsubsection{In relation to on-shell Weyl invariance}
We will briefly mention the ``on-shell" Weyl invariance studied in the literature in relation to the local renormalization group and its relation to the proof of ``a-theorem". The on-shell Weyl invariance is defined as a restricted Weyl invariance in the flat space-time limit. We consider the Weyl variation with $\Box \Omega = 0$ in the flat space-time limit, where $g_{\mu\nu} \to \eta_{\mu\nu}$ and $\Box \Omega = \eta_{\mu\nu} \partial^\mu \partial^\nu \Omega =0$. The on-shell Weyl invariance of the dimensionless geometric terms in \reff{S1} was the starting point of the argument in \cite{Komargodski:2011vj}. 

Obviously, the restricted Weyl invariance implies the on-shell Weyl invariance, but the converse is not necessarily true.
The infinitesimal on-shell Weyl invariance (i.e. $\Box \sigma = 0$ and $g_{\mu\nu} = \eta_{\mu\nu}$ {\it after} the Weyl variation), we have (only $S_{ij}(\phi) = T_{ijk}(\phi) = 0$ are imposed)
\begin{align}
a_{ij}(\phi) & = -\frac{1}{2} h_{ij}(\phi) \cr
b_{ijk}(\phi) & = -h_{k(i,j)}(\phi) +\frac{1}{2}h_{ij,k}(\phi)  
\end{align}
while 
$a(\phi)$, $c(\phi)$, $b(\phi)$ and $e_{i}(\phi)$, $f_{ij}(\phi)$, $h_{ij}(\phi)$, $c_{ijkl}(\phi)$ are arbitrary. 

However, the infinitesimal on-shell Weyl invariance is weaker than the finite on-shell Weyl invariance (i.e. $\Box \sigma = \partial^\mu \sigma \partial_\mu \sigma$) due to the lack of the Lie group structure (compare it with the restricted Weyl invariance in four dimensions studied above). For example, the finite on-shell Weyl invariance actually requires that $a(\phi)$ must be constant. It was crucial in section 2.1 that the consecutive restricted Weyl transformation is defined with the transformed metric $\tilde{\Box} \tilde{\Omega} = 0$, and the composition would not make sense if we restricted ourselves to the flat space-time only.

\subsubsection{Flat space-time energy-momentum tensor}
If a theory is invariant under the restricted Weyl invariant, and therefore, on-shell Weyl invariant, the trace of the flat space-time energy-momentum tensor is given by the d'Alembertian of a certain local composite field $O$.
\begin{align}
T^\mu_{\ \mu} = 2g^{\mu\nu} \frac{\delta S}{\delta g^{\mu\nu}}|_{g_{\mu\nu}= \eta_{\mu\nu}} = \Box O . \label{traceccc}
\end{align}
It should be contrasted with the rigid Weyl invariant case, where it is given by the divergence of the Virial current operator.
\begin{align}
T^\mu_{\ \mu} = 2g^{\mu\nu} \frac{\delta S}{\delta g^{\mu\nu}}|_{g_{\mu\nu}= \eta_{\mu\nu}} = \partial_\mu J^\mu .
\end{align}

The condition \eqref{traceccc} is nothing but that the theory living in flat space-time has a  conformal symmetry as a local non-gravitational field theory \cite{Polchinski:1987dy} (see \cite{Yu} for a review).
From these considerations, one can confirm that the dimensionless scalar action with two derivatives (\eqref{S1} with $g_{\mu\nu} = \eta_{\mu\nu}$) is conformally invariant in flat space-time, but the one with four derivatives (\eqref{S} with $g_{\mu\nu} = \eta_{\mu\nu}$) is generically only scale invariant but not conformal invraiant in flat space-time.

\subsection{Vector fields and rigid Weyl invariance with Euclidean signature}

So far our focus was the scalar field theories coupled with gravity. The introduction of the vector field $A_\mu$ with the Weyl weight zero add some interesting features because the rigid Weyl invariant (but not full or restricted Weyl invariant) action with two-derivative kinetic terms is possible. 

Let us begin with the case that the vector field is given by a gauge field. The introduction of the gauge field in the action is standard. We first replace all the derivatives in \eqref{S} with the gauge covariant derivative $\partial_\mu \phi^i \to D_\mu \phi^i = (\partial_\mu + iq A_\mu) \phi^i$.  In addition, we may introduce the gauge invariant kinetic term for the vector field
\begin{align}
S_{\text{gauge}} = \int d^4x \sqrt{|g|} \left(\frac{k}{4} F_{\mu\nu} F^{\mu\nu} + \frac{l_{ij}}{2} F^{\mu\nu} D_\mu \phi^i D_\nu \phi^j \right)  , \label{gaugeS}
\end{align}
where $F_{\mu\nu} = \partial_\mu A_\nu - \partial_\nu A_\mu$ is the field strength. The coefficients $k$ and $l_{ij}$ may depend on the scalar field $\phi$ with Weyl weight zero as long as they are gauge invariant.

The gauge potential $A_\mu$ has Weyl weight zero, and we can see that the gauge invariant kinetic terms are invariant under the full Weyl transformation by noting that no derivatives of the Weyl factor appears in \eqref{gaugeS}.

If we break the gauge invariance, there exist other terms that are not invariant under the full Weyl transformation.
\begin{align}
S_{\text{non-gauge}} &= \int d^4 x \sqrt{|g|} \left (m (D_\mu A^\mu)^2 + p R_{\mu\nu} A^\mu A^\nu + q R A_\mu A^\mu  \right) 
\end{align}
where $m,p,q$ may depend on $\phi$. 
The variation under the infinitesimal Weyl transformation is 
\begin{align}
\delta S_{\text{non-gauge}} &= \int d^4x \sqrt{|g|} (\partial^\mu \sigma) Z_\mu \cr
Z_\mu & = -4m A_\mu D^\nu A_\nu - (\partial_\mu (p A_\nu A^\nu) + 2 D_\nu (p A_\mu A^\nu) ) - 6 \partial_\mu (q A_\nu A^\nu)
\end{align}
Thus, we can see that $q R A_\mu A^\mu$ is restricted Weyl invariant but the other two terms are only rigid Weyl invariant.

Physically, the gauge non-invariant massless vector fields are realized in a theory of elasticity (emphasized by Riva-Cardy \cite{Riva:2005gd}). 
There the vector field $A_\mu$ represents the displacement vector.
In the flat Euclidean space, it is an example of scale invariant but not conformal field theory. In the curved background, it is only rigid Weyl invariant. The theory cannot be analytically continued to the Lorentzian signature due to the lack of reflection positivity. The coefficient $m$ of the rigid Weyl invariant (but not restricted or full Weyl invariant) term is known as the bulk modulus, and the curvature couplings of $p$ and $q$ were studied in \cite{ElShowk:2011gz} for the purpose of improvement of the energy-momentum tensor.

Furthermore, the non-gauge invariant interactions between the scalar and the vector fields are typically only rigid Weyl invariant. For instance, with the scalar field $\phi$ with Weyl weight zero, there are more non-gauge invariant terms like
\begin{align}
 S = \int d^4x \sqrt{|g|} (A^\mu D_\mu\phi) (D^\nu A_\nu) 
\end{align}
which  are only rigid Weyl invariant. A similar observation was made in \cite{Mack:1969rr, Gross:1970tb} in the context of scale invariant but non-conformal field theories in flat space-time.

\section{Conclusion}

In this paper, we showed that the dimensionless action \reff{S1} with scalar field of Weyl weight of $-1$ and arbitrary coefficients possesses restricted Weyl symmetry. In particular, we showed that the terms $R^2$, $R\phi^2$ and the minimal kinetic term for $\phi$ (i.e. $g^{\mu\nu} \partial_{\mu} \phi \partial_{\mu} \phi$), are not Weyl invariant but restricted Weyl invariant. To the best of our knowledge, the restricted Weyl symmetry of the dimensionless action \reff{S1} had not been previously emphasized in the literature. This is a non-trivial symmetry since the condition $\Box \Omega=0$ does not place serious limitations on $\Omega$ in contrast to rigid Weyl invariance where $\Omega$ must be a constant.  

One natural arena to explore restricted Weyl invariance is cosmology because Starobinsky's model of inflation \cite{Star} involves $R^2$ gravity.  Recently, the inflationary behaviour of $R^2$ gravity was studied in a conformal (Weyl) framework \cite{Lahanas}. The starting point of the authors' study was the action (Eq. 9 in their paper):
\beq
S=\int d^4x \sqrt{|g|} \Bigg(\dfrac{\alpha}{2} R^2 -\dfrac{s}{2} \Big(\dfrac{X^2}{6} R + (\nabla X)^2 \Big)  -\lambda X^4 \Bigg)   
\eeq{S3}
where $s=\pm 1$ and $X$ is a scalar field. As the authors point out, $R^2$ breaks the Weyl symmetry of the action. What we have shown is that though $R^2$ breaks Weyl symmetry it still possesses restricted Weyl symmetry. Action \reff{S3} above is precisely an example of an action which is not Weyl invariant but is restricted Weyl invariant. In the radiation-dominated era, the FLRW metric is restricted Weyl flat so that $R$ vanishes as well as the Weyl tensor. It is interesting to note that during this epoch, $R^2$ in \reff{S3} makes no contribution to the equations of motion i.e. the variation of $R^2$ with respect to the metric $g_{\mu\nu}$ is zero when $R=0$ in contrast to the variation of $R$.

The restricted Weyl invariance can be defined in any spacetime dimension, but we have realized that four space-time dimension is rather special.
Restricted Weyl transformations have the interesting property that consecutive restricted Weyl transformations generate another restricted Weyl transformation in four spacetime dimensions only. As a consequence, restricted Weyl transformations possess an inverse only in four dimensions.  
We also note that the gauge invariant vector field theory with two derivative kinetic terms studied in section 3.3 is only Weyl invariant in four dimensions. In other dimensions, it is only rigid Weyl invariant. We may speculate the origin of the four spacetime dimensional universe lies in the (restriced) Weyl invariance.

\section*{Acknowledgments}
A.E. acknowledges support from a discovery grant of the National
Science and Engineering Research Council of Canada (NSERC).  
Y.N. is supported by the World Premier International Research Center Initiative (WPI Initiative),

\begin{appendix}

\def\theequation{A.\arabic{equation}}
\setcounter{equation}{0}
\section{Transformation of kinetic term under Weyl transformation}

We now evaluate how the kinetic term for the scalar field in \reff{S1} transforms under the Weyl transformation \reff{Transform}. Using \reff{detg} and 
$g^{\mu\nu} \to \Omega^{-2}\, g^{\mu\nu}$ we obtain 
\begin{align}
&\sqrt{|g|} \,g^{\mu\nu} \,\nabla_{\mu}\phi \,\nabla_{\nu} \phi \cr
&\to \sqrt{|g|} \,g^{\mu\nu}\,\Omega^2\,\nabla_{\mu}(\phi\,\Omega^{-1})\, \nabla_{\nu} (\phi \,\Omega^{-1})\cr
& = \sqrt{|g|}\, \Big( \nabla_{\mu}\phi \,\nabla^{\mu}\phi - 2 \,\phi \,\Omega^{-1}\,\nabla_{\mu}\phi \,\nabla^{\mu} \Omega + \phi^2 \,\Omega^{-2}\,\nabla_{\mu}\Omega\,\nabla^{\mu}\Omega\Big)\cr 
& = \sqrt{|g|}\, \Big( \nabla_{\mu}\phi \,\nabla^{\mu}\phi - \nabla_{\mu} (\phi^2) \,\nabla^{\mu} (\ln \Omega) + \phi^2 \,\Omega^{-2}\,\nabla_{\mu}\Omega\,\nabla^{\mu} \Omega\Big)\cr
& = \sqrt{|g|}\, \Big( \nabla_{\mu}\phi \,\nabla^{\mu}\phi - \nabla_{\mu} (\phi^2 \,\nabla^{\mu} (\ln \Omega)) + \phi^2 \nabla_{\mu}\nabla^{\mu} (\ln\Omega) + \phi^2 \,\Omega^{-2}\,\nabla_{\mu}\Omega\,\nabla^{\mu} \Omega \Big)\cr
& = \sqrt{|g|}\, \Big( \nabla_{\mu}\phi \,\nabla^{\mu}\phi + \phi^2 \,\Omega^{-1}\,\Box \,\Omega - \nabla_{\mu} (\phi^2 \,\nabla^{\mu} (\ln \Omega)) \Big)\,.
\label{Kinetic2}\end{align}
The last term is a total derivative and makes no contribution to the action. We therefore see that if $\Box \Omega =0$, the kinetic term is invariant i.e. it is restricted Weyl invariant. We have therefore proved that the kinetic term possesses restricted Weyl symmetry.  

\def\theequation{B.\arabic{equation}}
\setcounter{equation}{0}

\section{$R_{\mu\nu}R^{\mu\nu}$ and $R_{\mu\nu\sigma \tau}R^{\mu\nu\sigma\tau}$ are restricted Weyl invariant terms}

The higher-derivative geometrical terms $R_{\mu\nu}R^{\mu\nu}$ and $R_{\mu\nu\sigma \tau}R^{\mu\nu\sigma\tau}$ can be expressed in terms of $\mathrm{Weyl}^2$ , $G$ and $R^2$ using \reff{Weyl} and \reff{Euler} i.e.
\begin{align}
R_{\mu\nu}R^{\mu\nu}&= \dfrac{1}{2} (\mathrm{Weyl}^2 -\mathrm{G}) + \dfrac{1}{3} R^2\,,\label{RR2}\\
R_{\mu\nu\sigma \tau}R^{\mu\nu\sigma\tau} &= 2\mathrm{Weyl}^2 -\mathrm{G} + \dfrac{1}{3} R^2\,.
\label{RR3}\end{align}
Under a Weyl transformation \reff{Transform}, $\sqrt{|g|} \,R^2$ transforms as \reff{R2}. The Gauss-Bonnet term transforms as \cite{Yu} 
\beq
\sqrt{|g|} \,\mathrm{G} \to \sqrt{|g|} \,\mathrm{G} + 4 \sqrt{|g|} \,\nabla^{\mu} (f_{\mu}) 
\eeq{Euler2}
where $f_{\mu}$ is given by 
\begin{align}
f_{\mu}= -R \,\Omega^{-1} \,\partial_{\mu} \,\Omega &+ 2 R_{\mu}^{\nu} \,\Omega^{-1} \,\partial_{\nu} \Omega -\partial_{\mu}(\Omega^{-2} \,\partial_{\nu} \Omega \,\partial^{\nu} \Omega)\cr 
&+ 
2\,\Omega^{-1} \,\partial_{\mu} \Omega (\Omega^{-1} \,\Box \Omega - \Omega^{-2} \,\partial_{\nu} \Omega \,\partial^{\nu} \Omega) + 2\,\Omega^{-3} \,\partial_{\nu} \Omega \,\partial^{\nu} \Omega \,\partial_{\mu} \Omega\,.
\label{fu} \end{align}
Therefore the Gauss-Bonnet term is Weyl invariant up to a total derivative. The Weyl tensor squared is a Weyl invariant i.e. under a Weyl tansformation we have
\beq
\sqrt{|g|} \,\mathrm{Weyl}^2 \to \sqrt{|g|} \,\mathrm{Weyl}^2 \,.
\eeq{Weyl2}
Using \reff{RR2}, \reff{RR3}, \reff{R2}, \reff{Euler2} and \reff{Weyl2} the tranformation properties of $R_{\mu\nu}R^{\mu\nu}$ and  \\$R_{\mu\nu\sigma \tau}R^{\mu\nu\sigma\tau}$ are  
\begin{align}
 \sqrt{|g|}\, R_{\mu\nu}R^{\mu\nu} &\to \sqrt{|g|} \,R_{\mu\nu}R^{\mu\nu} - 4 \,\sqrt{|g|} \,\Omega^{-1} \,R \,\Box \Omega \cr
&\qquad\qquad\qquad+ 12 \,\sqrt{|g|} \,\Omega^{-2} \,(\Box \Omega)^2 - 2 \sqrt{|g|}\, \nabla^{\mu} f_{\mu}\,,\label{R2R}\\
 \sqrt{|g|}\,R_{\mu\nu\sigma \tau}R^{\mu\nu\sigma\tau} &\to \sqrt{|g|} \,R_{\mu\nu\sigma \tau}R^{\mu\nu\sigma\tau}- 4 \,\sqrt{|g|} \,\Omega^{-1} \,R\, \Box \Omega\cr
&\qquad\qquad\qquad+ 12 \sqrt{|g|} \,\Omega^{-2} (\Box \Omega)^2 - 4 \sqrt{|g|} \,\nabla^{\mu} f_{\mu}\,.
\label{R3R}\end{align}
The above two terms are clearly not Weyl invariant. However, if $\Box \Omega=0$ then both terms above are invariant up to a total derivative. The total derivative makes a vanishing contribution to the action. We have therefore shown that $R_{\mu\nu}R^{\mu\nu}$ and $R_{\mu\nu\sigma \tau}R^{\mu\nu\sigma\tau}$ are restricted Weyl invariant terms just like $R^2$.  

\def\theequation{C.\arabic{equation}}
\setcounter{equation}{0}

\section{Variation of the action \reff{S}}

In this appendix we derive the variation \reff{DeltaS} of the action \reff{S} for the variation $\delta g_{\mu\nu}=-2 \sigma g_{\mu\nu}$ (or $\delta g^{\mu\nu} = 
2 \sigma g^{\mu\nu}$). The Weyl (conformal) weight of the scalar field is zero so that $\delta \phi=0$. We evaluate the variations term by term. In the process, we state and derive various formulas. The pertinent variations are 
\begin{align}
\delta \sqrt{|g|} &= -\dfrac{1}{2} \,\sqrt{|g|}\, \,g_{\mu\nu} \,\delta g^{\mu\nu} =-4 \,\sigma \,\sqrt{|g|}\,, \\\cr
\delta (\sqrt{|g|}\, \mathrm{Weyl}^2)&= \delta (\sqrt{|g|}\,) \mathrm{Weyl}^2 + \sqrt{|g|}\,\delta (\mathrm{Weyl}^2)\cr&
= -4 \sigma \sqrt{|g|}\, \mathrm{Weyl}^2 +4 \sigma \sqrt{|g|}\, \mathrm{Weyl}^2=0 \,,\\\cr
\delta (\sqrt{|g|}\, \mathrm{G}) &=-4 \sigma \sqrt{|g|}\, \mathrm{G} +4 \sigma \sqrt{|g|}\, \mathrm{G} \cr&\qquad\qquad +4 \sqrt{|g|}\,\nabla^{\mu}(R\, \partial_{\mu}\sigma - 2 R_{\mu}^{\nu} \partial_{\nu} \sigma)\cr&=4 \sqrt{|g|}\,\nabla^{\mu}(R\, \partial_{\mu}\sigma - 2 R_{\mu}^{\nu} \partial_{\nu} \sigma)\,,\\\cr
\delta (\sqrt{|g|}\, R^2)&= -4\sigma \sqrt{|g|}\, R^2 + \sqrt{|g|}\,2R\,\delta R=\sqrt{|g|}\, \,12 R\, \Box \sigma\cr
&\word{where} \delta R= 2 \sigma R +6 \,\Box \sigma \word{was used}\,,\\\cr
\delta (\sqrt{|g|}\, \partial^{\mu} R)&=2\sqrt{|g|}\, R\, \partial^{\mu}\sigma + 6\sqrt{|g|}\, \partial^{\mu}(\Box \sigma)\,,\\\cr
\delta (\sqrt{|g|}\, g^{\mu\nu} R)&= \sqrt{|g|}\, g^{\mu\nu} 6 \,\Box \sigma \,,\\\cr
\delta (\sqrt{|g|}\, G^{\mu\nu})&=\sqrt{|g|}\, g^{\mu\alpha}g^{\nu\beta} \delta G_{\alpha\beta}\cr&=\sqrt{|g|}\, g^{\mu\alpha}g^{\nu\beta}\delta(R_{\alpha\beta} -\dfrac{1}{2}g_{\alpha\beta} R)\cr&=\sqrt{|g|}\, (3\nabla^{\mu}\nabla^{\nu} \sigma -\nabla^{\nu}\nabla^{\mu} \sigma -2 g^{\mu\nu} \,\Box \sigma)\cr
\word{where} & \delta R_{\alpha\beta}= 3\nabla_{\alpha}\nabla_{\beta} \sigma -\nabla_{\beta}\nabla_{\alpha} \sigma + g_{\alpha\beta} \,\Box \sigma \word{was used}\,,\\\cr
\delta (\sqrt{|g|}\, \Box \phi^i\,\Box\phi^j)&=-2 \sqrt{|g|}\,\partial_{\mu}\sigma (\partial^{\mu}\phi^i \Box \phi^j +\partial^{\mu}\phi^j \,\Box \phi^i)\,,\\\cr
\delta (\sqrt{|g|}\, g^{\mu\nu}\,\Box\phi^k)&=-2 \sqrt{|g|}\,g^{\mu\nu} \partial_{\alpha}\sigma \partial^{\alpha}\phi^k \,,\\\cr
\delta (\sqrt{|g|}\, g^{\alpha\mu}g^{\nu\beta}\partial_{\mu}\phi^i\partial_{\alpha}\phi^j\partial_{\nu}\phi^k\partial_{\beta}\phi^l)&=0\,.
\end{align}

If one multiplies the above variations by their corresponding $\phi$ dependent coefficients, it is easy to verify that one obtains the variation \reff{DeltaS}. 

\end{appendix}


\begin{thebibliography}{99} 



\bibitem{Weyl}
H.~Weyl,
Berichte d. Preuss. Akad. d. Wissenschaften, 465 (1918).



\bibitem{O'Raifeartaigh:1996hf} 
  L.~O'Raifeartaigh, I.~Sachs and C.~Wiesendanger,
  DIAS-STP-96-06.

\bibitem{Osborn:1991gm} 
  H.~Osborn, Nucl.\ Phys.\ B {\bf 363}, 486 (1991).
\bibitem{Yu} Y. Nakayama, {\it Scale invariance vs conformal invariance}, arXiv:1302.0884. 
\bibitem{Komargodski:2011vj} 
  Z.~Komargodski and A.~Schwimmer, JHEP {\bf 1112}, 099 (2011) [arXiv:1107.3987 [hep-th]].
\bibitem{Hathrell} S. Hathrell, Ann. Phys. {\bf 139}, 136 (1982). 
\bibitem{Parker} L. Parker and D. Toms, {\it Quantum Field theory in Curved Spacetime}, Cambridge University Press, 2009.
\bibitem{BD} N. Birrell and P. Davies, {\it Quantum fields in curved space}, Cambridge University Press, 1982.
\bibitem{Nakayama:2012gu} Y.~Nakayama, Nucl.\ Phys.\ B {\bf 859}, 288 (2012) [arXiv:1201.3428 [hep-th]].
\bibitem{Carroll2} S. Carroll, {\it Spacetime and Geometry}, Addison Wesley, 2004.
\bibitem{BD2} N. Birrell and P. Davies, Phys. Rev. D {\bf 22}, 322 (1980).
\bibitem{FG}
C.~Feferman and C.~R.~Graham. 
Asterisque. (1985), 95-116.
\bibitem{GG}
C. R. Graham, R. Jenne, L. Mason and G. Sparling. 
Journal of London Mathematical Society. 46 (1992), 557-565.
\bibitem{G}
C.~R.~Graham. 
Journal of London Mathematical Society (2). 46 (1992), 566-576.
\bibitem{Sergei1} E. Elizalde, A.G. Zheksenaev, S.D. Odintsov and I.L. Shapiro, Class. Quant. Grav. {\bf 12}, 1385, (1995) [hep-th/9412061].
\bibitem{Sergei2}  E. Elizalde, A.G. Zheksenaev, S.D. Odintsov and I.L. Shapiro, Phys. Lett. B  {\bf 328}, 297, (1994) [hep-th/9402154].
\bibitem{Minas1} A. Padilla, D. Stefanyszyn and M. Tsoukalas, Phys. Rev. D {\bf 89}, 065009 (2014) [arXiv:1312.0975].
\bibitem{Riegert:1984kt} 
  R.~J.~Riegert,
  Phys.\ Lett.\ B {\bf 134}, 56 (1984).


\bibitem{Fradkin:1983tg} 
  E.~S.~Fradkin and A.~A.~Tseytlin,
  Phys.\ Lett.\ B {\bf 134}, 187 (1984).

\bibitem{PA}
S.~Paneitz, 
``A Quartic Conformally Covariant Differential Operator for Arbitrary Pseudo-Riemannian Manifolds", MIT preprint, 1983. Published posthumously in SIGMA 4 (2008), 036. [arxiv:0803.4331]


\bibitem{Osborn:2003vk} 
  H.~Osborn,
  Phys.\ Lett.\ B {\bf 561}, 174 (2003)
  [hep-th/0302119].


\bibitem{Polchinski:1987dy} 
  J.~Polchinski,
  Nucl.\ Phys.\ B {\bf 303}, 226 (1988).

\bibitem{Riva:2005gd} 
  V.~Riva and J.~L.~Cardy,
  Phys.\ Lett.\ B {\bf 622}, 339 (2005)
  [hep-th/0504197].

\bibitem{ElShowk:2011gz} 
  S.~El-Showk, Y.~Nakayama and S.~Rychkov,
  Nucl.\ Phys.\ B {\bf 848}, 578 (2011)
  [arXiv:1101.5385 [hep-th]].

\bibitem{Mack:1969rr} 
  G.~Mack and A.~Salam,
  Annals Phys.\  {\bf 53}, 174 (1969).
\bibitem{Gross:1970tb} 
  D.~J.~Gross and J.~Wess,
  Phys.\ Rev.\ D {\bf 2}, 753 (1970).
\bibitem{Star} A. Starobinsky, Phys. Lett. B {\bf 91}, 99 (1980); A. Starobinsky, Sov. Astron. Lett. {\bf 9}, 302 (1983).
\bibitem{Lahanas} A. Lahanas and K. Tamvakis, arXiv:1405.0828.
\end{thebibliography}
\end{document}